\documentclass[traditabstract]{aa} 

\usepackage{graphicx}
\usepackage{txfonts}
\usepackage{natbib}
\usepackage{longtable}
\usepackage{supertabular}

\begin{document}

\title{Asymmetric Heating of the HR~4796A Dust Ring \\Due to Pericenter Glow}

\author{Margaret M. Moerchen\inst{1,2}, Laura J. Churcher\inst{3}, Charles M. Telesco\inst{2}, Mark Wyatt\inst{3}, R. Scott Fisher\inst{4,5} \& Christopher~Packham\inst{2}}

\institute{
European Southern Observatory, Alonso de C\'ordova 3107, Casilla 19001, Vitacura, Santiago 19, Chile\\
\email{mmoerche@eso.org}
\and
Department of Astronomy, University of Florida, Gainesville, FL 32611, USA\\
\and
Institute of Astronomy, University of Cambridge, Madingley Road, Cambridge CB3 0HA, UK\\
\and
National Science Foundation, Division of Astronomical Sciences, 4201 Wilson Blvd., Suite 1045, Arlington, VA 22230, USA\\
\and
Gemini Observatory, Northern Operations Center, 670 N. AÕohoku Place, Hilo, HI 96720, USA
}

\authorrunning{Moerchen et al.}

%\begin{abstract}
\abstract
{
We have obtained new resolved images of the well-studied HR 4796A dust ring at 18 and 25 $\mu$m with the 8-meter Gemini telescopes.  These images confirm the previously observed spatial extent seen in mid-IR, near-IR, and optical images of the source. We detect brightness and temperature asymmetries such that dust on the NE side is both brighter and warmer than dust in the SW. We show that models of so-called pericenter glow account for these asymmetries, thus both confirming and extending our previous analyses.  In this scenario, the center of the dust ring is offset from the star due to gravitational perturbations of a body with an eccentric orbit that has induced a forced eccentricity on the dust particle orbits. Models with 2-$\mu$m silicate dust particles and a forced eccentricity of 0.06 simultaneously fit the observations at both wavelengths.  We also show that parameters used to characterize the thermal-emission properties of the disk can also account for the disk asymmetry observed in shorter-wavelength scattered-light images.} {}{}{}{}

%Dust of this size is below the blowout radius, and a relatively high mass loss rate (2 x 10$^{22}$ g yr$^{-1}$) would be required to sustain this population in a steady state.
%\end{abstract}


\maketitle

\section{Introduction}

HR~4796A is an A0V star with the highest fractional IR luminosity ($L_{IR}$/$L_{\star}$= 5 x $10^{-3}$) yet discovered among debris disks. 
Its recently revised distance estimate is 73~pc  \citep{van07} (updated from 67~pc [Perryman et al. 1997\nocite{per97}]), and it has an M-star T~Tauri companion with a separation of 7.7'' \citep{jur93}.  Its age of 8$\pm$2~Myr \citep{sta95} (based on the lithium abundance of HR~4796B) places the disk of HR~4796A in the very interesting early phase of chaotic disk evolution dominated by collisions (e.g., Kenyon \& Bromley 2005\nocite{ken05}).

Based on the dust temperature, \citet{jur93} first tentatively suggested 40~AU (now 44~AU) as the main location of dust responsible for the HR~4796A excess thermal emission.  Indeed, the mid-IR discovery images of the source at CTIO and Keck revealed a highly inclined ringlike disk with a dust distribution peaking near 70~AU (now 76~AU) \citep{jay98, koe98}.  Follow-up coronagraphic near-IR images with $HST$/NICMOS tightly constrained the width of the ring to $\sim$17~AU (now $\sim$18.5~AU), with the dust population severely depleted both interior and exterior to this region \citep{sch99}.

Further imaging observations of the thermal emission confirmed a strong peak in the dust density at 70~AU \citep{tel00, wah05}.  \citet{tel00} also noted a 1.8-$\sigma$ brightness asymmetry at 18~$\mu$m in which the northeast side of the disk is brighter than the southwest side.  In a companion paper, \citet{wya99} demonstrated that such an asymmetry could arise from the phenomenon of pericenter glow.  A second body in the system on an eccentric orbit about the star may cause pericenter glow when, through secular perturbations, it effectively shifts the center of the dust ring away from the star and closer to the apastron of the perturbing companion's orbit.  The side of the ring shifted closer to the star becomes relatively warmer and more luminous \citep{der98}, giving rise to the ``glow''.  At present, while several works have acknowledged the existence of a brightness asymmetry (e.g., Wahhaj et al. 2005, Debes et al. 2008)\nocite{wah05, deb08}, no alternative explanations for the brightness asymmetry have been proposed.

We have obtained mid-IR images of HR~4796A with the Michelle and T-ReCS cameras at the Gemini North and Gemini South telescopes, respectively.  These permit us to confirm unambiguously the brightness asymmetry, which was previously characterized at only the 1.8-$\sigma$ level of significance. By imaging thermally emitting dust at two wavelengths, we also examine the spatial distribution of dust temperatures within the resolved disk, and we conclude that the dust population in the NE ansa of the disk is not only brighter, but also warmer, than the dust in the SW ansa.  With these new data and with the consideration of shorter-wavelength images of scattered light, we re-examine the possibility that these observed asymmetric characteristics are caused by pericenter glow.  
%%The similar morphology of HR~4796A and Fomalhaut is also noted: both dust rings are well-defined in radial extent and possess disk centers that are offset from their host star.  A planetary companion associated with Fomalhaut has recently been confirmed through multi-epoch imaging \citep{kal08}.  HR~4796A may also host a planetary-mass object that could be detected with current or near-future imaging facilities, although the distance of the HR~4796A system is nearly ten times greater than that of Fomalhaut.

\section{Mid-IR images of HR 4796A}

\begin{figure}[!t]
\begin{centering}
\includegraphics[width=\columnwidth]{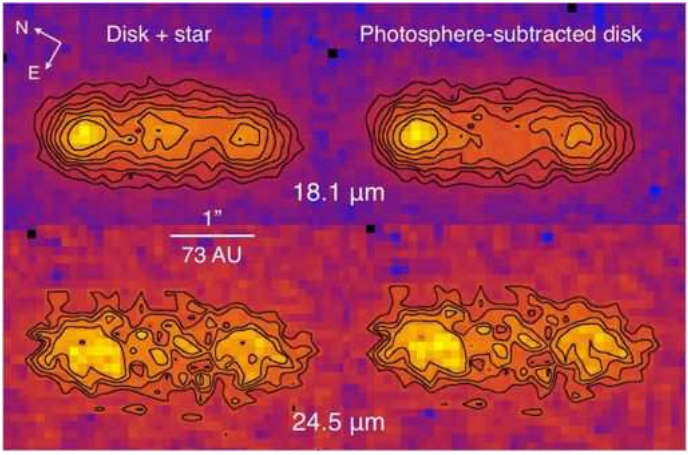}
\caption[18.1-$\mu$m and 24.5-$\mu$m  image contours. ]{Top: 18.1-$\mu$m  image contours drawn at 2-$\sigma$ intervals starting at 5$\sigma$ (background~$\sigma$~=~0.295 mJy/pixel).  Bottom: 24.5-$\mu$m  image contours drawn at 1-$\sigma$ intervals starting at 3-$\sigma$ (background~$\sigma$~=~1.463 mJy/pixel). Note that the 18.1-$\mu$m image was taken with Michelle (plate scale: 0.1005''/pixel) and the 24.5-$\mu$m image was taken with T-ReCS (plate scale: 0.089''/pixel).  The images have been scaled to show the same area in square arcseconds, and both images are photosphere-subtracted.\label{fig:4796xconvimages}}
\end{centering}
\end{figure}

We observed HR~4796A with T-ReCS at Gemini South (using the Qb filter) in May 2004 and with Michelle at Gemini North (using the Qa filter) in April 2005.  The central wavelengths for these two filters are 18.1~$\mu$m for Qa and  24.5~$\mu$m for Qb.  A PSF and flux standard star were observed close in time to the target object for both sets of images.  We reduced the images with an IDL reduction package developed at the University of Florida, and we performed photometric measurements with the Starlink GAIA program.  The images in both bandpasses show that the disk of HR~4796A is clearly resolved, in the form of an elongated disk $\sim$3'' in extent with two prominent peaks in brightness (Figure \ref{fig:4796xconvimages}).

Table~\ref{tab:4796photom} lists results of the sky-subtracted aperture photometry of HR~4796A at 18.1~$\mu$m and 24.5~$\mu$m.  The uncertainties reflect only measurement errors.  Photometric uncertainties are driven primarily by variable sky transmission throughout the night, and lacking multiple standard star measurements, we adopt typical photometric uncertainties of 15\% for both filters.

\begin{table}[!h]
	\caption{Photometry of the HR~4796A Disk}\label{tab:4796photom}
	\begin{tabular}{cccc}
 \hline
$\lambda_c$ & Total Flux Density & Photosphere & Disk Flux Density \\
$[\mu$m] & [mJy] & [mJy] & [mJy] \\
 \hline
18.1 & 1106 $\pm$ 7 & 48 & 1058 $\pm$ 7 \\

24.5 & 3307 $\pm$ 47 & 33 & 3274  $\pm$ 47 \\
\hline
	\end{tabular}
	
\begin{flushleft}
\small{Notes-- Uncertainties listed in this table are measurement uncertainties.  Photometric uncertainties are assumed to be 15\% for both bandpasses. 
	}
\end{flushleft}
\end{table}

The photospheric contribution in the bandpasses of the imaging observations are based on a 10.8-$\mu$m value derived by \citet{jur98} by extrapolating the $K$-band (2.2~$\mu$m) flux density using a $\nu^{1.88}$ power law as given by the \citet{kur79} model atmosphere with T~=~9500~K and log~$g$~=~4.0.  Photospheric values at longer wavelengths were estimated with the assumption that HR~4796A and Vega have the same 10-to-18~$\mu$m flux ratio, since they are both A0V stars.  The photospheric contribution to the total flux density of the source is 4\% at 18.3~$\mu$m and 1\% at 24.5~$\mu$m.

The PSF was scaled such that the integrated flux matched the photosphere level, and its center was shifted to the position assumed for the star in the target source images.  The maximum intensities of the scaled PSF and of the disk (prior to the photosphere subtraction) at the stellar position are, respectively, 1.4~mJy/pixel and 5.0~mJy/pixel at 18.1~$\mu$m and 0.3~mJy/pixel and 9.8~mJy/pixel at 24.5~$\mu$m.  For the 18.3-$\mu$m image, the adopted stellar position was the centroid peak of the 2-pixel-smoothed central lobe of emission (which we assume arises from the presence of the star) between the NE and SW ansae, where the peaks were also determined from a centroid measurement of the 2-pixel-smoothed image.  The stellar centroid peak position in the 18.1-$\mu$m image differs from the actual midpoint between the NE and SW peaks of emission by $<$1 pixel, which is insignificant as this is approximately the level of certainty in the determination of the peak location.  One pixel in the 18.1-$\mu$m image corresponds to 0.1'', or $\sim$7~AU.  For the 24.5-$\mu$m image, the stellar position is not obvious, and so the midpoint between the two centroid peaks of the NE and SW ansae was adopted as the stellar position.  The distance from the NE peak to the SW peak in the 2-pixel-smoothed image is 17.3~pixels (1.74'') at 18.1~$\mu$m and 18.2~pixels (1.62'') at 24.5~$\mu$m.

The scaled PSF image in each bandpass was then subtracted from the corresponding image of HR~4796A.  
The total photosphere-subtracted flux density of the source was measured at each wavelength, and the results are given in Table~\ref{tab:4796photom}.

\section{Disk asymmetry}\label{sec:diskmeas}

The global dust color temperature is 102~$\pm$~11~K, based on the 18.1-$\mu$m/24.5$\mu$m flux density ratio determined from aperture photometry of the entire photosphere-subtracted disk.  This temperature is consistent with the initial color temperature estimated by \citet{jur93} based on $IRAS$ photometry (T$\sim$110~K).  The color temperature of the dust as a function of disk radius was also calculated using the approach described below.

\begin{figure}[]
\includegraphics[width=\columnwidth]{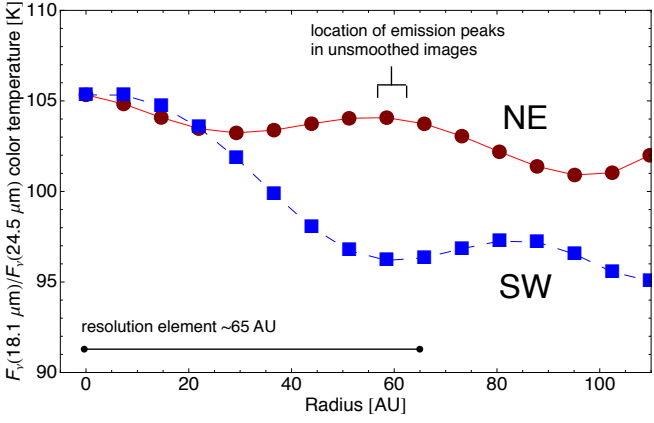}
\caption[Color temperature profile of the HR~4796A disk based on photometric measurements at 18.1~$\mu$m and 24.5~$\mu$m. ]{Color temperature profile of the HR~4796A disk based on photometric measurements at 18.1~$\mu$m and 24.5~$\mu$m (images ``cross-convolved'' to achieve the same spatial resolution).  The NE profile is represented by circles, and the SW profile is represented by squares.  Combined measurement uncertainties from both bandpasses are smaller than the size of the data points. The temperatures at the location of the peak dust density in the ring (labeled) are the key values to note.}\label{fig:4796temp}
\end{figure}

\begin{figure}[!h]
\includegraphics[width=\columnwidth]{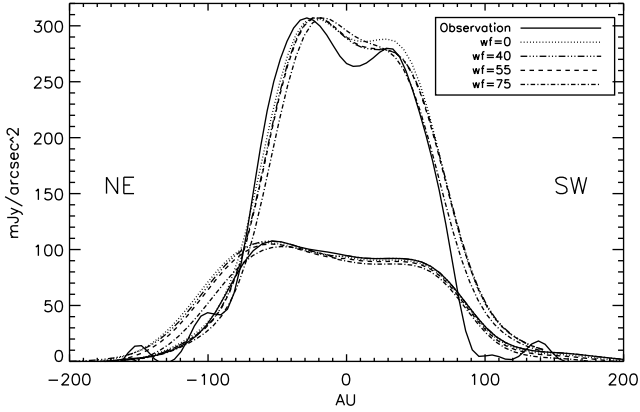}
\caption[Brightness profiles of the HR~4796A observations and model at 18.1~$\mu$m (lower curves) and 24.5~$\mu$m (upper curves).]{Brightness profiles of the HR~4796A observations and model at 18.1~$\mu$m (lower curves) and 24.5~$\mu$m (upper curves).  The brightness profile for each observation or disk model was generated by summing the central 3'' along the long axis of the disk.  Model curves are shown for four different angles of pericenter, and we have concluded that a pericenter angle of zero provides the best fit. \label{fig:linecuts}}
\end{figure}

%\begin{figure}[!h]
%\includegraphics[width=\columnwidth]{brightnessprofiles_labels.eps}
%\caption[Brightness profile of the HR~4796A disk at 18.1~$\mu$m (squares) and 24.5~$\mu$m (circles).]{Brightness profiles of the photosphere-subtracted HR~4796A disk at 18.1~$\mu$m (squares) and 24.5~$\mu$m (circles).  Measurement uncertainties are within the size of the data points.  Photometric uncertainties are assumed to be 15\% of the measured flux density. \label{fig:4796bright}}
%\end{figure}

The position angle (PA), east of north, is defined as the angle between the vertical image axis and the line connecting the two central peaks of emission of the two ansae: 28.1$^{\circ}$ in the 18.1-$\mu$m image and 29.9$^{\circ}$ in the 24.5-$\mu$m image.  We estimate an uncertainty for the PA of 0.9 degrees, half the difference between the values measured in the stacked image at each wavelength.  We adopt a PA value of 29$^{\circ}$, the average of the two values.  For convenient display, the images of HR~4796A were rotated counter-clockwise by an angle of 61$^{\circ}$, to orient the disk plane parallel to the x-axis.  
%%  (90$^\circ$ minus the PA of the disk E of N)

\textit{Quantifying the asymmetric structure--}
The asymmetric structure is quantified in both the observations and in the model distributions (\S\ref{sec:modeling}) by finding the brightness peak for each ansal ``lobe'' and its offset from the star.  We calculated the following values for each image: 

\begin{itemize}
\item $D_{mean}$, the mean offset of the lobes' peak brightness from the center, in pixels
\item $\frac{\Delta D}{D_{mean}}$, the difference between the distance of the lobe brightness peaks ($D_{NE}-D_{SW}$), divided by the mean offset of the brightness peaks from the center, given as a percentage value
\item $F_{mean}$, the mean peak brightness of the lobes
\item $\frac{\Delta F}{F_{mean}}$, the difference in lobe peak brightness, ($F_{NE}-F_{SW}$), divided by the mean peak brightness of the lobe ($F_{mean}$), given as a percentage value
\end{itemize}

The level of the brightness asymmetry, $\Delta F/F_{mean}$, is of primary interest, and it is 15.3$~\pm$~2.6\% at 18.1~$\mu$m and 13.0$~\pm$~3.8\% at 24.5~$\mu$m.  The uncertainty in $\Delta F/F_{mean}$ was calculated by taking a circular symmetrical fit to the observations (i.e., repeating the modeling but imposing zero forced eccentricity) and then using Monte Carlo methods to add noise to the model at the observed level (0.007~mJy/pixel at 18.3~$\mu$m, 0.135~mJy/pixel at 24.5~$\mu$m) to work out the asymmetry that would have been detected purely due to noise.

\textit{Assessing the temperature asymmetry--} 
After removal of the photospheric contribution, the angular resolution of each image was degraded to achieve the same resolution in both images for accurate spatial comparison in the color temperature calculation.  Using the $gauss$ routine in IRAF, each of the target images was convolved with a Gaussian profile having the same FWHM as the PSF from the other bandpass.  The PSF FWHM was estimated by a Gaussian fit to the azimuthally averaged profile.  The FWHM values were 5.19~pixels (0.52'') in the 18.1-$\mu$m image and 8.08~pixels (0.72'') in the 24.5-$\mu$m image.  The PSF images themselves were convolved in the same manner, which confirmed that the resulting resolution (assessed with FWHM measurements) was identical for both images.  The final resolution of the cross-convolved images is $\sim$0.9'', which corresponds to $\sim$65~AU for the source distance of 73~pc. 

The rotated images were cropped to a swath of vertical width $\Delta$y$\sim$3''.  The pixel values in these images were summed along the y-axis, resulting in a one-dimensional brightness profile for the disk at each wavelength.  
From these brightness profiles, a color profile and a color temperature profile (Figure~\ref{fig:4796temp}) were constructed to illustrate variations in temperature along the extent of the disk.   As a test of the robustness of the temperature estimates, the 24.5-$\mu$m brightness profile was shifted by $\pm$1 pixel ($\sim$6~AU) along the x-axis relative to the 18.1-$\mu$m brightness profile before the two brightness profiles were combined to determine the color profile.  The two resulting color profiles that incorporated these offsets yielded color temperature profiles that did not differ from the results presented in Figure~\ref{fig:4796temp} within 60~AU of the star by more than 2~K.

The NE and SW profiles are overplotted on the same radius scale (along the x-axis) to better show differences in temperature at the same distances from the central star.  We see no significant variation in color temperature within a radius of 15~AU from the star.  While that result is expected for such a narrow annulus, this result is in fact attributable principally to the degradation of resolution that we carried out to make the temperature estimates.  This effect is also seen in the separation of the two ansae peaks, which decreased for both images following the cross-convolution, from 1.74'' to 1.1'' at 18.1~$\mu$m, and from 1.62'' to 1.58'' at 24.5~$\mu$m (where the lesser effect for the 24.5~$\mu$m image is likely due to being convolved by a smaller kernel).

There is a clear temperature asymmetry, with the dust in the NE side of the disk having significantly higher color temperatures at greater than 30~AU from the star and in particular at the location of the emission peaks corresponding to the two ansae of the dust ring.  The temperature difference peaks at $\sim$9~K near 60~AU.  The uncertainties due to measurement error in the temperature estimates are less than 2~K.  The color temperature uncertainties that incorporate photometric uncertainties are $\pm$8~K, but in this analysis we are most interested in the relative temperatures of the two ansae and are therefore most concerned with uncertainties within the images and not absolute photometric uncertainties.

%% POSSIBLY OMIT THE FOLLOWING PARAGRAPH.
%% We have examined the effect of over- and under-subtraction of the photospheric flux density from the disk image as follows.  We scaled the PSF to yield the same integrated flux density as the photosphere and measured the peak flux density value of this scaled PSF, 1.4~mJy/pixel at 18.1~$\mu$m and 0.34~mJy/pixel at 24.5~$\mu$m.  We calculated the photospheric flux density that would be yielded if this peak surface brightness were sustained across the width of the Airy disk of the PSF ($\sim$10~pixels).  Using the brightness profiles of the photosphere-subtracted images, the color at the center of the disk profile is $\sim$0.33.  If the two MIR brightness levels are assumed to have the overestimated uncertainties described above (i.e., such that the peak brightness value occurs throughout the Airy disk), then the color would have an associated uncertainty of 0.05.  However, even if the photosphere-scaled PSF were over- or under-subtracted, the effect on the estimated color would be negligible beyond the bounds of the Airy disk, which corresponds to $\sim$35~AU in disk radius.  Indeed, a significant color asymmetry is still apparent at radii greater than 35~AU, and so our conclusion of a temperature asymmetry remains unaffected.

\section{Pericenter glow modeling}
\label{sec:modeling}

\subsection{Pericenter glow scenario and prior work}

One motivation for observing debris disks is the potential to reveal as-yet unseen planetary companions through the disk morphology.  A massive orbiting body can gravitationally perturb the orbits of smaller bodies like dust particles over long-period (``secular'') timescales, and these perturbations may be observed in the global characteristics of the disk.  For example, a planet on an eccentric orbit eventually imposes a forced eccentricity on the dust particles, which results in an offset of the disk's center of symmetry from the host star.  While this offset may be observable directly, other manifestations of the offset may be more prominent.  In particular, if the dust population is azimuthally homogenous in size distribution and composition, the side of the disk (the pericenter side) that is offset toward the star will experience enhanced heating, increasing both its temperature and brightness.  The resulting asymmetry in brightness between the apocenter and pericenter sides of the disk is referred to as pericenter glow \citep{der98}.

The asymmetric morphology of the HR~4796A disk observed in previous thermal emission images at 18~$\mu$m was well approximated by models that invoked pericenter glow \citep{wya99, tel00}, but the statistical significance of the asymmetry in those earlier images was relatively low.  We note that the brightness asymmetry was also previously detected at 10.8~$\mu$m \citep{tel00}, but the much greater prominence of the starlight at those wavelengths makes such shorter wavelength observations less useful for our analysis.  With new 18.1-$\mu$m images and the addition of images at 24.5~$\mu$m in this dataset, we have re-examined this hypothesis by using the same models to reconstruct the observed thermal emission and color temperature profiles.

\subsection{Thermal emission reconstruction}

\textit{Model inputs--} The physics of pericenter glow and the associated modeling of it are outlined in detail by \citet{wya99}, and here we review only the key parameters involved in that analysis.  
%These models have also been used by Grogan et al. 1997 and Dermott et al. 1994 to study the zodiacal cloud.
The model for the disk's density distribution of dust was generated with three distinct inputs: the physical structure of the disk, the combination of the optical properties and size distribution of the dust particle population, and the orientation of the disk along the line of sight. We recall that the mass of the perturbing body itself does not come into play and is not constrained by this approach.  The physical structure of the disk is modelled by a radial density distribution of dust with defining parameters as follows: $a_{min}$ and $a_{max}$, the inner and outer semimajor axes of the dust ring (which are roughly equivalent to the inner and outer radii), $\gamma$, the power-law index of the semimajor axis distribution, $e_f$, the forced eccentricity, and $\sigma_{tot}$, the total surface area of the dust particles in the disk.   The forced eccentricity is that imposed on all particles in the disk by the planet's secular perturbations \citep{mur99}; as such, the planet's orbit is not affected by its own perturbations.  The eccentricities of particle orbits are the vector sum of this forced eccentricity and their own proper eccentricity.  To first order, the disk remains circular but is just offset.  The total surface area of the dust is effectively a scaling factor that can be adjusted to approximate the overall brightness level of the dust emission.  While the intrinsic disk offset and asymmetry (i.e., if viewed face-on) are determined by the forced eccentricity $e_f$, the measured offset and asymmetry are determined by the additional variables of disk inclination, $i$, and the angle of pericenter, ${\omega}_f$.  For example, the forced eccentricity required to replicate a given measured asymmetry increases as the angle of pericenter deviates from being perpendicular to the line of sight. 

The optical properties of dust were calculated using Mie theory, Rayleigh-Gans theory, and geometric optics \citep{li97,aug99,boh83} for astronomical silicate spheres the diameter of which ($D_{typ}$=2~$\mu$m) was constrained by a simultaneous fit to the images at both wavelengths.  The resulting dielectric constant was 6.7x$10^{-5}$, and the absorption coefficients were 0.54 and 0.30 at 18.1 and 24.5~$\mu$m, respectively.
While a single particle size is used to represent a particle population that has a range of sizes and composition, it reasonably encapsulates key mid-IR emission characteristics of the disk.
We adopted values for the stellar luminosity and effective temperature of 21~L$_{\odot}$ and 9500~K, respectively (e.g., Debes et al., 2008).

%% a 2-$\mu$m  of density 2500~kg~m$^{-3}$, which is smaller than the blowout diameter for the system: $D_{blowout}~=~7.7~\mu$m for a particle with zero porosity and $\beta$~=~0.5.   The 2-$\mu$m size provides a 

%*** I calculated a blowout diameter (for $\beta$=0.5) of 7.7 microns, but I note in previous documents you listed a value of 19.12 microns.  I used L$_{\star}$=21L$_{\odot}$ and M$_{\star}$=2.5M$_{\odot}$ -- are these the same values you used? ***

The output from the disk radial distribution model and the calculation of the dust optical properties were combined with the specified orientation to calculate the line-of-sight brightness and generate images of the model disk.  The model images at each wavelength were rotated to the same position angle as the observed disk, and each image was convolved by a PSF kernel with the profile width of the observed PSF reference star in the same filter.  In the generation of color temperature profiles for the model data, the images were also convolved by the profile width corresponding to the other filter to match the treatment of the observational data (\S\ref{sec:diskmeas}).
%%Then, the observed and modeled images were both smoothed by a Gaussian function with $\sigma$~=~5~pixels to eliminate spikiness prior to the assessment of disk asymmetry.

\begin{figure}[!h]
\includegraphics[width=\columnwidth]{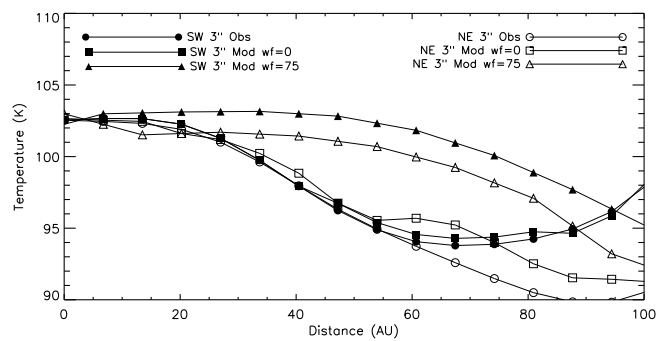}
\caption[Color temperature profiles of the HR~4796A observations and models.]{Color temperature profiles of the HR~4796A observations and models.  Colors were calculated by summing the flux within a 3''-wide swath along the long axis of the disk for each image.  Model curves are shown for two different angles of pericenter, and we have concluded that a pericenter angle of zero provides the best fit. \label{fig:tempcuts}}
\end{figure}

\textit{Optimizing the model--}
With the images of the model disk, we generated brightness profiles and color temperature profiles in the same way that we did with the observational data (\S\ref{sec:diskmeas}).  We compared the observed and model profiles, and we adjusted the model parameters until the model disk yielded a good approximation to the observations.  First, we varied the inner and outer disk radii and the total disk surface area until the brightness profile width and magnitude were well approximated.  We then adjusted the forced eccentricity until the brightness asymmetry level was matched for a fixed value of the angle of pericenter.  The fit was repeated for several values of the angle of pericenter to assess the corresponding forced eccentricity necessary to reproduce the same asymmetry (Table~\ref{tab:asymmetry}).  The forced eccentricity and the angle of pericenter define the magnitude of the radial offset asymmetry in the disk model, which in turn defines the brightness asymmetry and the dust color temperature asymmetry.  The initial assessment of how changing these parameters affected the resulting brightness profiles was performed by chi-squared minimization on the brightness profiles and images.  The brightness profiles of the observations and the model disk with varying values for the angle of pericenter are shown in Figure~\ref{fig:linecuts}, and the color temperature profiles are shown in Figure~\ref{fig:tempcuts}.  The quantitative measurements used to assess the asymmetry in the model and in the observed disk, as described above, are summarized in Table~\ref{tab:asymmetry}.

\begin{table*}
\caption{Asymmetry parameters for observations and models\label{tab:asymmetry}}
\centering
\begin{tabular}{lccccccc}
\hline\hline
& & \multicolumn{2}{c}{18.1} & & \multicolumn{2}{c}{24.5} & \\ 
\cline{3-4}  \cline{6-7} 
 & $e_f$ & $\frac{\Delta D}{D_{mean}}$ [\%] & $\frac{\Delta F}{F_{mean}}$ [\%]  & & $\frac{\Delta D}{D_{mean}}$ [\%] & $\frac{\Delta F}{F_{mean}}$ [\%]  & reduced $\chi^2$  \\
\hline
observed & -- & 0 & 15.28  & & 0 & 13.00 & --\\
model $\omega_f=0^\circ$ & 0.06 & -11.71 & 15.26 & & -6.11 & 13.10 & 2.17 \\
model $\omega_f=40^\circ$ & 0.10 & -10.81 & 15.31 & & -7.01 & 13.30 & 2.31 \\
model $\omega_f=55^\circ$ & 0.13 & -11.18 & 15.35 & & -7.15 & 13.05 & 2.21 \\
model $\omega_f=75^\circ$ & 0.30 & -14.50 & 10.55 & & -7.39 & 12.70 & 2.26 \\
\hline

\end{tabular}
\tablefoot{
All models were simultaneously fitted to 18.1- and 24.5-$\mu$m brightness profile linecuts of 0.5'' and share the following parameter values: $a_{min}$=70~AU, $a_{max}$=84~AU, $\gamma$=-1.5, $D_{typ}$=2~$\mu$m, and $i$=14.1$^{\circ}$.  Also recall that the value $\frac{\Delta D}{D_{mean}}$ for the observations is 0 because we defined the stellar position at 24.5~$\mu$m as the midpoint between the brightness peaks of the two ansae (with the measured difference between the midpoint and the stellar peak at 18.1~$\mu$m to be $<$1 pixel.)}
\end{table*}

\textit{Results of the modelling--}
We have determined that a forced eccentricity of 0.06 imposed on a ring of dust with $D_{typ}$=~2~$\mu$m spanning 74--84~AU can reproduce the thermal IR images, fitting the observed brightness asymmetry well at both 18.1 and 24.5~$\mu$m.  The disk density power-law exponent was set to -1.5. The same result for the characteristic particle size was inferred previously in order to simultaneously fit 10-$\mu$m and 18-$\mu$m images of HR~4796A \citep{wya99, tel00}.  We have estimated the uncertainty in the forced eccentricity to be 0.01, which is the change required to effect a 1-$\sigma$ change in the brightness asymmetry of the model at 18.3~$\mu$m (the more limiting case).  The angle of pericenter was 0$\pm$30$^\circ$ (perpendicular to the line of sight), and the disk inclination was 14.1$^{\circ}$ \citep{sch09}.  We note that these results differ from those found by \citet{wya99} and \citet{tel00} ($e_f$=0.02, $\omega_f$=$75^\circ$) likely because of the different levels of observed asymmetry, which are nonetheless within 3$\sigma$ of one another (5.1~$\pm$~3.2\% in the previous work, 15.3~$\pm$~2.6\% in this work).  

We have also used the model to quantify the limits set by these data on the presence of hot dust inside the ring. We compared the photometry of the photosphere-subtracted images with that of the model within an aperture radius equal to the size of the PSF FWHM (0.52" at 18.3~$\mu$m, 0.72" at 24.5~$\mu$m), and then we repeated the measurement with the same aperture on the models, where an unresolved flux component could be added at the position of the star. Excluding calibration uncertainties, the measured flux density within a 0.52"-radius aperture on the 18.3-$\mu$m photosphere-subtracted image was 289~$\pm$~3 mJy. The same measurement on the model with no added unresolved flux component was 290~mJy. Therefore, the maximum unresolved contribution we can add to the model that is 3-$\sigma$ consistent with the observations is 8~mJy. Repeating the same procedure at 24.5 um with an aperture of radius 0.72", we measured a flux density of 1290~$\pm$~12 mJy for the photosphere-subtracted images and 1289~mJy for the model. Therefore, the 3-$\sigma$ upper limit of 1326~mJy for this aperture allows the presence of an unresolved component at a level of 37~mJy. However, these measurements do not account for the assumed 15\% calibration uncertainty. Since a change in calibration would also result in a corresponding change to the model (to fit the peaks at the correct level), it is instructive to consider what would have been derived with a revised calibration factor.  In the extreme situation that the calibration factor was 45\% higher, the flux measured in the previously described apertures at 18.1 and 24.5~$\mu$m respectively would have been 438~$\pm$~4~mJy and 1881~$\pm$~17~mJy, with corresponding model fluxes of 290*1.45~=~421~mJy and 1289*1.45~=~1869~mJy, and so a flux excess of 17~$\pm$~4 and 12~$\pm$~17~mJy. Therefore, a 3-$\sigma$ deviation in both calibration and statistical uncertainty puts the limits on the unresolved flux at $<$29~mJy and $<$63~mJy at 18.1 and 24.5~$\mu$m. \citet{wah05} estimated an exozodiacal contribution of 87~$\pm$~22~mJy at 24.5~$\mu$m, which is consistent to within 3$\sigma$ with the upper limit we derive. 

%41.3 mJy within 18um aperture, 22.7 in 25um aperture

\subsection{Scattered light comparison}

A brightness asymmetry has also been detected at wavelengths that are dominated by light scattered by the dust disk.  Images obtained by \citet{sch99} at 1.1~$\mu$m were suggestive of a brightness imbalance favoring the NE that was later noted to be at the 10--15\% level \citep{sch01}.  The disk ansae were again compared in the analysis of data sets taken with $HST$ NICMOS at 1.71--2.22~$\mu$m, but the NE ansa was not consistently brighter in all bandpasses \citep{deb08}.  Most recently, $HST$ STIS images at 0.57~$\mu$m show that the SW ansa is 0.74~$\pm$~0.07 times as bright as the NE ansa, as measured in ``lobes'' within $\pm$20$^\circ$ of the major axis and between 0.8'' and 1.5'' from the star. \citet{sch09} measured an offset asymmetry, $(D_{NE}-D_{SW})/D_{mean}~=-3.7\%$, such that the center of the disk is offset from the star by 1.4~$\pm$~0.4~AU (19~$\pm$~6~mas) in the plane perpendicular to the line of sight.  
%%According to the inverse-square falloff in brightness, this difference in distance from the star alone should result in the NE side being 1.075 times (1.036$^2$) as bright as the SW side.  Therefore, 17.5\% (25\% minus 7.5\%) of the brightness asymmetry seen in the optical is left to be explained.

Thus, another test for our pericenter glow model is whether it can replicate the observations in scattered light.  To examine this, we modified the model's particle properties to correctly calculate the scattered light fluxes, using a combination of Mie theory, the Rayleigh-Gans approximation, and geometric optics, which yielded a mean albedo of 0.05.  We assumed that the same particle population (in this case, modeled as a first approximation by 2-$\mu$m silicate spheres) is responsible for both the thermal emission and the scattered light reflection.  We adopted a Henyey-Greenstein phase function to model the non-isotropic scattering properties of the small grains \citep{aug99}, with the form
\begin{equation}\label{eq:phasefxn}
f(\alpha) = \frac{1 - g^2}{1 - g^2 -2gcos(\alpha)}
\end{equation}
where $\alpha$ is the scattering angle to the line of sight and $g$ is the asymmetry parameter.  The parameter $g$ indicates the relative level of front or back scattering, such that $g=1$ for 100\% forward scattering, $g=-1$ for 100\% backward scattering, and $g=0$ for isotropic scattering.  We adopted the value $g=0.16$ as determined from the optical $STIS$ images \citep{deb08, sch09}, which is consistent within 2$\sigma$ of the values for $g$ determined by \citet{deb08} for the near-IR $NICMOS$ images.

To check that our model produced realistic flux levels, we compared them to those measured by \citet{sch09} at a wavelength of 0.57~$\mu$m.  Due to the presence of coronagraphic spikes in the image, the flux density measurement had to be corrected for the missing area, yielding a total flux 9.4$\pm$0.8~mJy.  The total flux in our model scattered light image is 10.8~mJy.  As a secondary check, we used the formula given by \citet{wei99} to predict the surface brightness of a disk in scattered light (in mJy arcsec$^{-2}$)
\begin{equation}\label{eq:diskflux}
S = \frac{F\tau\omega}{4\pi\phi^2}
\end{equation}
where $F$ is the flux received from the star in mJy, $\tau$ is the optical depth of the scattering material, $\omega$ is the albedo (from Mie theory, where $Q_{sca}$=0.08), and the factor $\tau\omega$ can be approximated by $L_{disk}/L_{\star}$ if $\omega=1$.
The expected flux density calculated with this relation agrees with the flux density values both from observations and from our model to within 20\% at 0.57, 1.1, and 1.6~$\mu$m.

%\begin{table}
%\caption{Comparison of scattered light flux measurements\label{tab:diskflux}}
%\begin{tabular}{lccc}
%\tableline\tableline
%Wavelength & F[mJy] prediction & F[mJy] observed & F[mJy] model   \\
%\tableline
%0.57~$\mu$m & 9.26 & 9.4 & 10.8     \\
%1.1~$\mu$m & 6.4 & 7.6 & 5.4  \\
%1.6~$\mu$m & 6.4 & 7.4 & 7.6  \\
%\tableline
%\end{tabular}
%\end{table}

We also replicated the lobe asymmetry measurement by \citet{sch09} with our disk model.  Within the same apertures (within $\pm$20$^\circ$ of the major axis and between 0.8'' and 1.5'' from the star), we determine a SW/NE brightness ratio of 0.92
%$~\pm$~0.06 
at 0.57~$\mu$m and 0.93
%$~\pm$~0.05 
at 1.1~$\mu$m.  Our model values are therefore consistent within 3$\sigma$ with the \citet{sch09} value of 0.74$~\pm~$0.07, and are also consistent with a simple 1/$r^2$ attenuation of starlight due to the disk ansae being at different distances from the star.  Our model also replicates the radial offset asymmetry at a level of 3.3\%, which is consistent with the \citet{sch09} value of 3.7~$\pm$~1.1\%.

\citet{sch09} acknowledge that dynamical perturbations such as those we propose may be responsible for part of the brightness asymmetry, also suggesting that a segregation of the particle population or an azimuthal density variation in a homogenous dust ring could produce the asymmetry.  However, whether such a model could also reproduce the observed temperature asymmetry has yet to be tested.  Additionally, we note that particles in the disk should be moving slightly faster at pericenter, relative to their velocity at apocenter, and the collisional timescale at pericenter should be relatively shorter in turn.  Therefore, the population of small particles that scatter near-IR light may be enhanced due to the increased collision rate near pericenter, thus yielding a density variation in the dust ring and a greater brightness imbalance in scattered light.  There is the potential to further develop the pericenter glow model with such an enhanced particle population, which may help to determine whether factors beyond dynamical perturbations must be invoked to explain the observed morphology.  Such refinements to our model would bear greater consideration especially if the large near-IR brightness asymmetry is confirmed.

\section{Discussion and conclusions}

Using new mid-IR images at 18~and 25~$\mu$m, we have confirmed the previously observed brightness asymmetry discovered by \citet{tel00} in the HR~4796A disk to a high level of statistical significance.
We have determined that a model disk with forced eccentricity of 0.06 that is dynamically induced by a companion can replicate that brightness asymmetry.  After incorporating the dust scattering properties, we can also reproduce, to within 3$\sigma$, the level of brightness asymmetry seen in near-IR scattered-light images .  We believe that near-IR imaging in additional bandpasses will further constrain the level of the asymmetry observed in scattered light, and that such constraints will assist in confirming whether the pericenter glow model is sufficient to explain the disk asymmetry observed throughout the spectrum. 

The model disk is comprised of particles 2~$\mu$m in diameter ($D_{typ}$), in agreement with previous analysis of the disk \citep{wya99, tel00}, and the particles are assumed to be spherical silicates.  Dust particles smaller than $\sim$8~$\mu$m in diameter should be blown out roughly within an orbital timescale ($\sim$400~yr).  In reality the dust population is not this simple, but we have adopted this value to characterize the overall population.  We must consider that the particles may have a more complex set of optical properties that allows them to be heated to the temperatures we infer and to remain in the system without being rapidly blown out.  More complex dust properties (e.g., Li \& Lunine 2003\nocite{li03}) could be incorporated into the pericenter glow model in future iterations.

Finally, we recall that a key result for a system like HR~4796A that exhibits pericenter glow is the implied presence of a perturbing companion.  In fact, the principal perturbing body may be the nearby M-star companion HR~4796B.  However, the orbital parameters of HR~4796B are unknown and may not be commensurate with its being the primary perturber.  The sharp inner edge of the ring points to the possibility that the perturbing object responsible for the asymmetry is also truncating the inner edge, as has been suggested for Fomalhaut \citep{kal05a, qui06}.  In this case, the putative planet responsible would be expected to lie just interior to the ring.  For example, a Jupiter-mass planet would be expected to orbit at $\sim$60~AU with an eccentricity of $\sim$0.06 (maximum projected separation 0.87''), but it could be closer to the dust ring if it were more massive.  Given the ambiguity in potential perturbers and the fact that the level of forced eccentricity induced by the companion does not constrain its mass, the continued search for the putative companion is warranted, if not with the current generation of telescopes then with the next, such as the space-based $JWST$ or the ground-based TMT or E-ELT facilities.  \citet{kal08} have demonstrated a successful direct planet detection that was pursued following similar evidence of a forced stellar offset, as in the case of Fomalhaut, which is nearer and brighter than HR~4796A.

\begin{acknowledgements}

MMM gratefully acknowledges fellowship support from the Michelson Science Center. This work was performed [in part] under contract with 
JPL funded by NASA through the Michelson Fellowship Program. JPL is managed for NASA by the California Institute of Technology.  LJC is grateful for the support of an STFC studentship.  Observations were obtained at the Gemini Observatory, operated by AURA, Inc., under agreement with the NSF on behalf of the Gemini partnership: NSF (US), PPARC (UK), NRC (Canada), CONICYT (Chile), ARC (Australia), CNPq (Brazil), and CONICET (Argentina).

\end{acknowledgements}


%% in parentheses, after the keyword, but they will not be verified.

\bibliographystyle{apj}
\bibliography{../moerchen_bibtex}

\end{document}